\documentstyle[prc,aps]{revtex}
\input{psfig}
\begin{document}
\draft
\title{Measurement of neutron capture on $^{48}$Ca at thermal and
       thermonuclear energies}
\author{H. Beer}
\address{Institut f\"ur Kernphysik, Forschungszentrum Karlsruhe,
                           P.~O.~Box 3640, D-76021 Karlsruhe, Germany}
\author{C. Coceva}
\address{ENEA, Via Don Fiammelli 2, I-40128 Bologna, Italy}
\author{P. V. Sedyshev, and Yu. P. Popov}
\address{Frank Laboratory of Neutron Physics, JINR, 
               141980 Dubna, Moscow Region, Russia}
\author{H. Herndl, R. Hofinger, P. Mohr, and H. Oberhummer}
\address{Institut f\"ur Kernphysik, Wiedner Hauptstr.~8-10, TU Wien,
                  A-1040 Vienna, Austria}
\date{\today}
\maketitle
\begin{abstract}
At the Karlsruhe pulsed 3.75\,MV Van de Graaff accelerator the
thermonuclear $^{48}$Ca(n,$\gamma$)$^{49}$Ca(8.72\,min) cross section was
measured by the fast cyclic activation technique via the 
3084.5\,keV $\gamma$-ray line of the $^{49}$Ca-decay.
Samples of CaCO$_3$ enriched in $^{48}$Ca by 77.87\,\%
were irradiated between two gold foils which served as capture standards.
The capture cross-section was measured at the neutron energies 25, 151,
176, and 218\,keV, respectively.
Additionally, the thermal capture cross-section was measured at the
reactor BR1 in Mol, Belgium, via the prompt and decay $\gamma$-ray lines
using the same target material.
The $^{48}$Ca(n,$\gamma$)$^{49}$Ca cross-section in the thermonuclear
and thermal energy range has been calculated using the direct-capture
model combined with folding potentials. The potential strengths 
are adjusted to the scattering length and the binding energies of the 
final states in $^{49}$Ca. The small coherent elastic
cross section of $^{48}$Ca+n is
explained through the nuclear Ramsauer effect.
Spectroscopic factors of $^{49}$Ca have been extracted
from the thermal capture cross-section with better accuracy than from a
recent (d,p) experiment. Within the uncertainties both results are in
agreement.
The non-resonant thermal and thermonuclear experimental data for this reaction
can be reproduced using the direct-capture model.
A possible interference with a resonant contribution is discussed.
The neutron spectroscopic factors of $^{49}$Ca determined from
shell-model calculations are compared with the values
extracted from the experimental cross sections for
$^{48}$Ca(d,p)$^{49}$Ca and $^{48}$Ca(n,$\gamma$)$^{49}$Ca.
\end{abstract}



\section{Introduction}\label{s1}
For a long time it has been known that the solar-system abundances of elements
heavier than iron have been produced by
neutron-capture reactions~\cite{bur57}. However, neutron capture
is also of relevance for abundances of isotopes
lighter than iron especially for neutron-rich
isotopes, even though the bulk of these elements
has been synthesized by charged-particle induced reactions.
The attempts to understand neutron-induced nucleosynthesis require as
important ingredients the knowledge of neutron-capture
rates. The influence of shell effects on neutron capture is
one of the most interesting aspects of neutron capture, especially
since neutron capture in the vicinity of magic numbers is often
a bottleneck in neutron-induced nucleosynthesis. This is the case
also in neutron capture on neutron-rich isotopes
close to the magic proton and neutron numbers $Z=20$ and $N=28$, i.e., 
in the vicinity of the double-magic nucleus $^{48}$Ca. The
neutron capture for nuclei in this mass region is of relevance
for the Ca-Ti abundance anomalies occurring in certain
primitive meteorites~\cite{san82,zie85,woe93}.

Neutron-capture on the double-magic nucleus
$^{48}$Ca at thermal and thermonuclear energies
is also of interest from the viewpoint of nuclear
structure. This nucleus has an excellent closed-shell
structure and the level density of the compound-nucleus $^{49}$Ca at
neutron-separation energy is very low~\cite{uoz94}. In fact, no
strong resonances have been observed for $^{49}$Ca
in (d,p)-reactions or beta-delayed neutron decay below 172\,keV. 
The two small resonances at 19.3\,keV and 106.9\,keV
have only been observed in neutron capture~\cite{car85,car87}, but have
not been found in $\beta$-delayed neutron decay~\cite{zie80,car82}
and in the (d,p)-reaction~\cite{uoz94,met75}. They are probably $d$-wave resonances
with a very small spectroscopic strength. Therefore,
it can be expected that the resonant compound-nucleus
contribution for $^{48}$Ca(n,$\gamma$)$^{49}$Ca
can be neglected and that the non-resonant direct capture
is the dominant reaction mechanism. This has also been
verified in a previous direct-capture calculation of
$^{48}$Ca(n,$\gamma$)$^{49}$Ca~\cite{kra95}.

Neutron capture on $^{48}$Ca(n,$\gamma$)$^{49}$Ca has been
measured previously at thermal~\cite{cra71} and thermonuclear
energies~\cite{car87,kae85}. This data basis appeared to us not yet
sufficient. At thermal neutron energy the measurement has a relative
large uncertainty of 15\%. The measurement was carried out with a
sample of low $^{48}$Ca enrichment using a Moxon--Rae detector.
The thermonuclear time--of--flight measurement \cite{car87} was
only sensitive to resonance capture and in the activation measurement
\cite{kae85} only two values were determined,
using a Maxwell neutron spectrum of kT=25\,keV
and a very broad spectrum from a few keV to 170\,keV (average energy: 
97\,keV). 
\section{Experimental setup and results for thermal neutron capture}\label{s2}
\par
The capture cross-section at thermal energy was obtained by measuring 
with a Ge-crystal the intensity of gamma-rays emitted with known
probability after neutron capture in $^{48}$Ca.
More precisely, our experimental method is based on the following
considerations.
In the measured spectrum of capture gamma-rays, the number of counts
$A(E_\gamma)$ in a peak corresponding to a given energy
$E_\gamma$ can be expressed as a product of four terms:
\begin{equation}
A( E_\gamma )  =  N_{\rm inc}  p_{\rm a}  f ( E_\gamma )  \varepsilon
( E_\gamma ) \quad ,
\label{a1}
\end{equation}
where $N_{\rm inc}$ is the total number of neutrons incident on
the sample, $p_{\rm a}$ is the capture probability (per incident
neutron) in $^{48}$Ca, $f(E_\gamma)$ is the {\it a priori}
known emission probability (per capture event) of the gamma-ray
under consideration, and $\varepsilon(E_\gamma)$ is its
probability of being detected.
The dependence of $A(E_\gamma)$ on the sought cross section
$\sigma_\gamma $ is contained in the probability $p_{\rm a}$.
For a thin sample (as in our case), the effect of capture in
other nuclides, and of scattering can be considered as a correction
for which a very precise knowledge of the relevant cross sections
is not needed. In practice, $p_{\rm a}$ can be assumed to be a known
function of $n\sigma_\gamma$, where $n$ is the thickness
(atoms per barn) of $^{48}$Ca.

The product $N_{\rm inc}\,\varepsilon(E_\gamma)$ in Eq.~(\ref{a1})
must be deduced from a separate measurement
of the gamma spectrum from the standard
reaction $^{35}$Cl(n,$\gamma)$, using the same experimental set-up.
In this calibration run, with
reference again to Eq.~(\ref{a1}), the probability $p_{\rm a}$ can be calculated
from the known capture cross-section of $^{35}$Cl and from the
total cross-section of all sample components. Values of $f$
are known for several calibration lines in the range of interest
\cite{Coc96}.
Summarizing, thanks to supplementary information on the emission
probability $f$ and on the product $N_{\rm inc}\,\varepsilon$,
we can state that some selected peak areas of the
$^{48}$Ca(n,$\gamma)$ spectrum are known functions of the
capture cross-section.

We should like to point out that the merit of this method lies in
its intrinsic ability of picking out only those capture events
occurring in a given nuclide of the sample.
In our experiment advantage was taken of the precisely known \cite{Ch69}
emission probabilities $f$(3084.54\,keV)=$0.921\pm0.010$ and
$f$(4071.9\,keV)=$0.070\pm0.007$. These gamma-rays are emitted
after $\beta^-$ decay of $^{49}$Ca with a half life of
523\,s.
Moreover,
we observed prompt gamma-rays in $^{49}$Ca, i.e., the
ground state transition at 5142\,keV and the two-step cascade
at 3121 and 2023\,keV. It is known \cite{Ar69} that the intensity
sum of the primary ground state transition and of the cascade through
the 2023\,keV level is very close to 100\,\%. Therefore, we could
use also these prompt transitions in $^{49}$Ca to deduce the
capture cross-section.

\begin{figure}[t]
\unitlength1cm
\centerline{\psfig{file=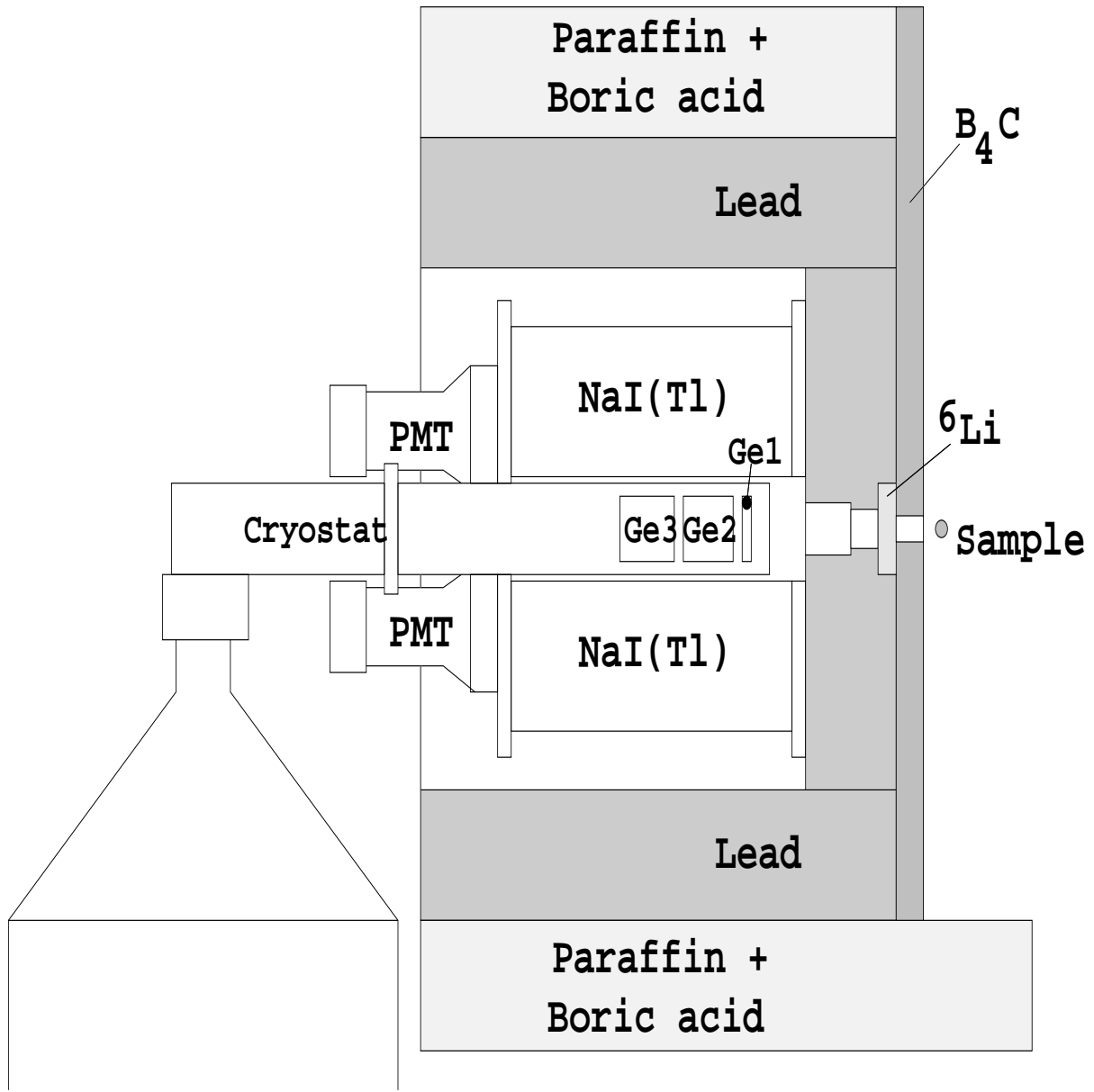,width=16cm,height=16cm,angle=270}}
\vspace{1cm}
\caption{\label{ff1} Scheme of experimental setup of the thermal measurement}
\end{figure}

The experiment was carried out at the BR1 reactor of the
"Studiecentrum voor kernenergie", Mol (Belgium), at an experimental
channel supplying a thermal neutron flux of about
$5 \times 10^5$/(cm$^2$\,s) on the sample.
The experimental set-up is schematically shown in Fig.~\ref{ff1}.
The sample consisted of about 0.1\,g of Calcium carbonate
(CaCO$_3$) containing 77.87\,\% enriched $^{48}$Ca (Tables~\ref{tt1} and~\ref{th2}).
Carbonate powder was enclosed in a cylindrical teflon container 
with 0.6\,cm inner diameter.

As shown in Fig.~\ref{ff1}, all detectors were shielded with particular
care by means of Boron carbide and metallic $^6$Li. In fact
the rate of scattered neutrons from the sample and its container
was two orders of magnitude higher than the capture rate in
$^{48}$Ca.
Gamma-rays emitted by the sample were detected in a $130$\,cm$^3$
coaxial Ge crystal (labelled Ge2 in Fig.~\ref{ff1}), completely
surrounded by a 30\,cm\,$\times$\,30\,cm NaI(Tl) scintillator, by
a planar Ge crystal on the front side, and by a coaxial Ge crystal
on the rear. All three Ge crystals were mounted in the same
cryostat, housed in an 8\,cm diameter through-hole along the axis of
the NaI(Tl)  cylinder.

A very effective suppression of escape peaks and Compton tails was
obtained by rejecting all pulses from the central Ge crystal in
coincidence with at least one of the other detectors.
The energy resolution (FWHM) at 3\,MeV was 2.9\,keV.
In the calibration run, CaCO$_3$ was replaced by a 61.835\,mg sample
of Carbon hexachloride (C$_2$Cl$_6$) in a similar teflon container.
Here knowledge of the capture cross-section of $^{35}$Cl
($\sigma_\gamma=(43.6\pm0.4)$\,barn) and of the total cross-section
of the sample components Cl and C \cite{Mug81} allow a very
precise calculation of the capture probability $p_{\rm a}$. Emission
probabilities $f(E_\gamma)$ of 24 measured lines between
0.5 and 8.6\,MeV are known within 3\,\% \cite{Coc96}.

Using the above data, the product $N_{\rm inc}\,\varepsilon$ for five
energies of $^{49}$Ca and
$^{49}$Sc lines (Fig.~ref{ff2}) was calculated from the curve
\begin{equation}
N_{\rm inc}\,\varepsilon =a_1\,+\,a_2E^{-a_3}_\gamma\,\exp(-E_\gamma/
a_4) \quad,
\label{a2}
\end{equation}
where the four parameters $a_1,\ldots,a_4$ were obtained by best
fitting the 24 calibration points of the Chlorine spectrum. It
was demonstrated \cite{Ow91} that expression (\ref{a2}) can reproduce
very accurately the energy dependence of Ge-detector efficiencies
over a wide range. In fact, in our fit we obtained a normalised
chi-square value $\chi^2=1.1$.
The measurement with the $^{48}$Ca sample lasted 35 hours,
while the calibration with $^{35}$Cl took 15 hours.

In conclusion, from the measured counts under the full-energy peaks
of three lines of $^{49}$Ca and of two lines of $^{49}$Sc, the
following consistent set of data was obtained:
$f$(5142\,keV)\,=\,$0.74\pm0.03$; $f$(3121\,keV)\,=\,$f$(2023\,keV)\,=\,$0.23\pm0.01$,
and $p_{\rm a}\,=\,(1.632\pm0.032) \times 10^{-3}$.
Finally, the capture cross-section for $^{48}$Ca deduced from
the above $p_{\rm a}$
value is $$\sigma_\gamma=(0.982\,\pm\,0.046) \, {\rm barn}\quad .$$
The quoted error is comprehensive of a systematic
2.4\,\% uncertainty in the enrichment of the $^{48}$Ca sample.
The results are summarized in Table~\ref{th2}.
\section{Fast Cyclic Activation Technique and thermonuclear 
              capture cross-sections}\label{s3}
The thermonuclear measurements have been carried out at the Karlsruhe 
pulsed 3.75\,MV Van de Graaff accelerator. The technique of fast cyclic
activation has been described in detail in previous publications 
\cite{bee94,bee95}. To gain statistics
the frequent repetition of the irradiation and
activity counting procedure which characterizes a cycle
is essential.
The time constants for each cycle which are chosen shorter than the
fluctuations of the neutron beam and comparable or shorter than the decay
rate $\lambda$ of the measured isotope $^{49}$Ca
are the irradiation time $t_{\rm b}$, the counting time $t_{\rm c}$, the
waiting time $t_{\rm w}$ (the time to switch from the irradiation to the
counting phase) and the total time T=$t_{\rm b}$+$t_{\rm w}$+$t_{\rm
c}$+$t'_{\rm w}$
($t'_{\rm w}$ the time to switch from the counting to the irradiation phase).
In the actual $^{48}$Ca measurements the runs were partly carried out with
$t_{\rm b}$=49.58\,s, $t_{\rm c}$=49.24\,s, and T=100\,s, partly with $t_{\rm
b}$=99.58\,s, $t_{\rm c}$=99.24\,s, and T=200\,s.
The waiting time is in both cases $t_{\rm w}$=0.42\,s.

The accumulated number of counts from a total of $n$ cycles,
$C=\sum_{i=1}^n C_i$, where $C_i$,
the counts after the i-th cycle, are calculated for a chosen irradiation
time, $t_{\rm b}$, which is short
enough compared with the fluctuations of the neutron flux, is~\cite{bee94}
\begin{equation}
\label{eq1}
C  = \epsilon_{\gamma}K_{\gamma}f_{\gamma}\lambda^{-1}[1-\exp(-\lambda
t_{\rm c})]
\exp(-\lambda t_{\rm w})
 \frac{1-\exp(-\lambda t_{\rm b})}{1-\exp(-\lambda T)} N \sigma
{[1-f_{\rm b} \exp(-\lambda T)]}
 \sum_{i=1}^n \Phi_i
 \end{equation}
with
\begin{displaymath}
f_{\rm b}  =  \frac{\sum_{i=1}^n \Phi_i \exp[-(n-i)\lambda T]}{
\sum_{i=1}^n \Phi_i} \quad .
\end{displaymath}
The following additional quantities have been defined;
$\epsilon_\gamma$: Ge-efficiency, $K_\gamma$:
$\gamma$-ray absorption, $f_\gamma$: $\gamma$-ray intensity per decay,
$N$: the number of target
nuclei, $\sigma$: the capture cross-section, $\Phi_i$: the neutron flux
in the i-th cycle. The
quantity $f_{\rm b}$ is calculated from the registered flux history of a $^6$Li
glass monitor.

The efficiency determination of the 35\,\% HPGe-detector (2\,keV resolution
at 1.332\,MeV) has been reported elsewhere~\cite{bee92}. The $\gamma$-ray
absorption was calculated using tables published by Storm and
Israel~\cite{sto70} and Veigele~\cite{vei73}.
The half-lives and the $\gamma$-ray intensities per decay of $^{49}$Ca 
and $^{198}$Au are given in Table~\ref{tt1}.

The activities of nuclides with half lives of several hours
to days, i.e., the activity of $^{198}$Au, is additionally counted after
the end of the cyclic activation consisting of $n$ cycles using
\begin{eqnarray}
\label{eq2}
C_n=\epsilon_\gamma K_\gamma f_\gamma \lambda^{-1} [1-\exp(-\lambda T_{\rm M})]
\exp(-\lambda T_{\rm W})
[1-\exp(-\lambda t_{\rm b})] N \sigma f_{\rm b} \sum_{i=1}^n \Phi_i \quad.
\end{eqnarray}
Here $T_{\rm M}$ is the measuring time of the Ge-detector and $T_{\rm W}$
the time
elapsed between the
end of cyclic activation and begin of the new data acquisition.

Eqs.~(\ref{eq1}) and (\ref{eq2}), respectively, contain the
quantities $\sigma$
and the total neutron flux $\sum_{i=1}^n \Phi_i$. The 
unknown capture cross-section of $^{48}$Ca is measured relative
to the well-known standard cross section of
$^{197}$Au 
\cite{mac82,rat88}. As the $^{48}$CaCO$_3$ sample to be investigated is
characterized by a finite thickness
it is necessary to sandwich the sample by two comparatively thin gold
foils for the determination
of the effective neutron flux at sample position. The activities of these
gold foils were counted also
individually after termination of the cyclic activation. The effective
count rate of gold was obtained from these individual rates as well as
from the accumulated gold count rate during the cyclic activation run.
Therefore, the effective neutron flux at sample position was determined
in two ways by way of the gold activation according to the Eqs.~(\ref{eq1})
and (\ref{eq2}).
Using Eq.~(\ref{eq1}) has the advantage that
saturation
effects in the gold activity for irradiations over several days are
avoided~\cite{bee94}.

\begin{figure}[t]
\centerline{\psfig{file=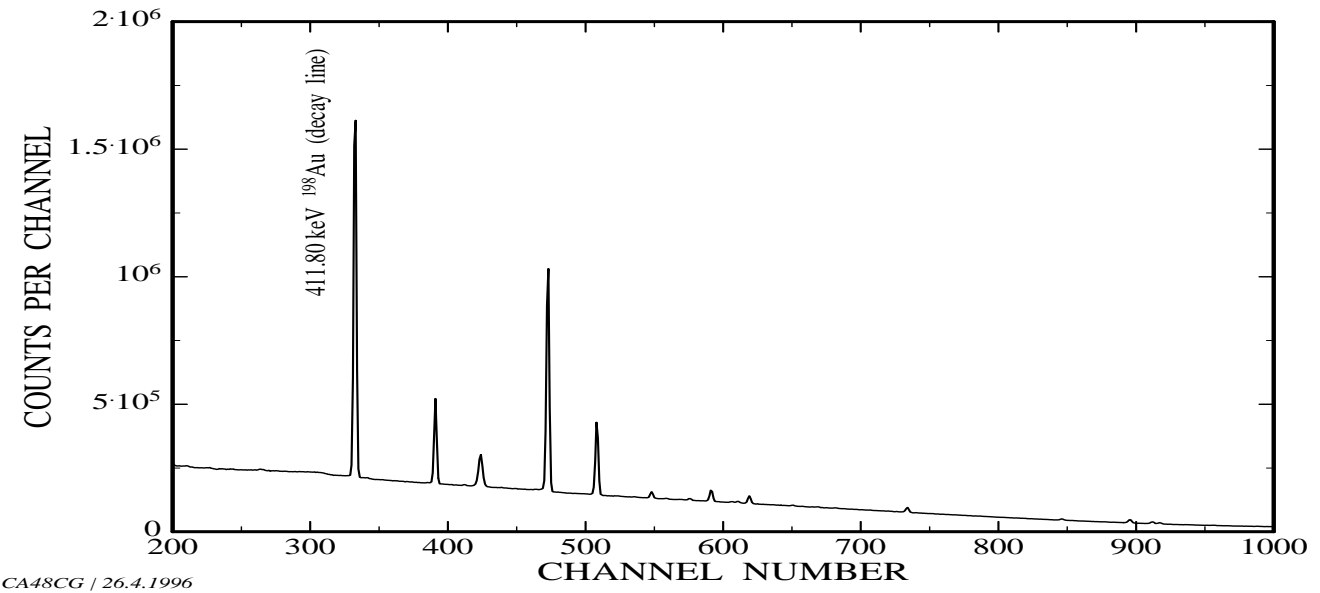,width=18cm,height=8cm,angle=270}}
\vspace{1cm}
\centerline{\psfig{file=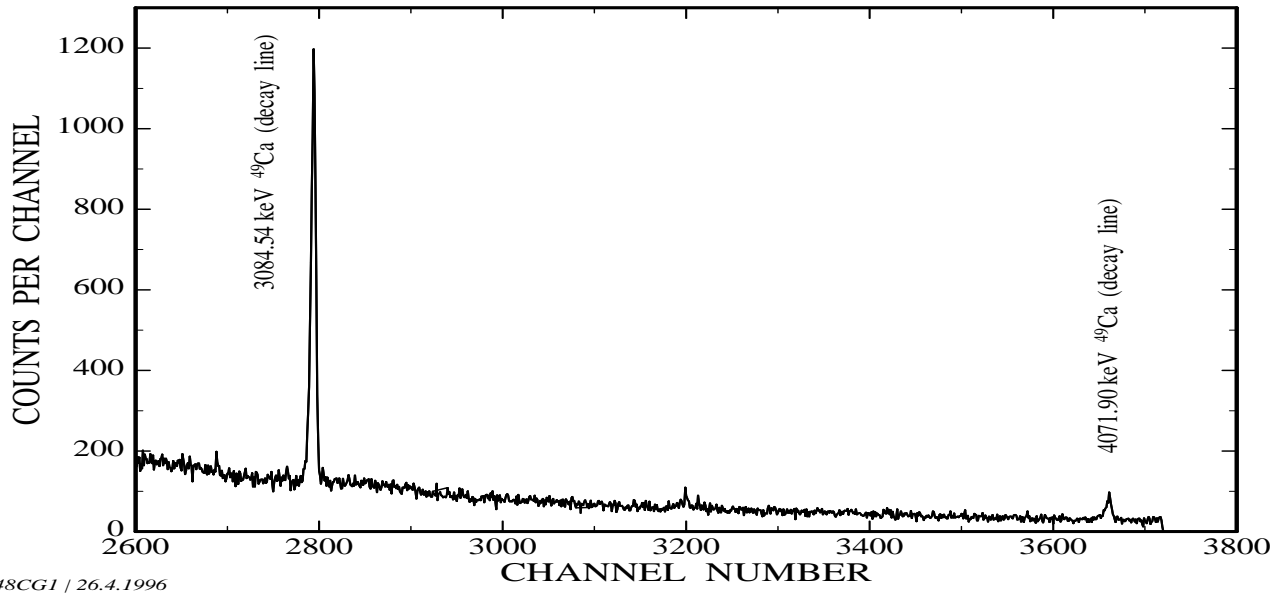,width=18cm,height=8cm,angle=270}}
\vspace{1cm}
\caption{\label{ff3} Accumulated intensities of the
$^{49}$Ca (below) and $^{198}$Au (above) $\gamma$-ray decay lines
from the activation with a 50.40\,mg CaCO$_3$ sample sandwiched
by two Au foils using a neutron spectrum
with a mean energy of 176\,keV}
\end{figure}

The neutron spectrum at the
neutron energy 25\,keV was generated using proton energies close to the 
reaction threshold. This condition produces a kinematically collimated 
neutron spectrum resembling a Maxwellian spectrum at a temperature of 
25\,keV. The spectra at 151,
176, and 218\,keV were generated using thin Li-targets (2.5$\,\mu$m).
The required proton energy
conditions and the neutron spectra integrated over the solid angle of the
sample were determined in time-of-flight (TOF) measurements
before the actual activation runs using the accelerator in pulsed
mode. The measurements were carried out under the same conditions as 
previously reported for the $^{36}$S(n,$\gamma$)$^{37}$S experiment 
\cite{bee95}.
Table~\ref{tt2} gives a survey of the sample weights and the measured
$^{48}$Ca capture cross-sections. The $^{48}$CaCO$_3$ sample of 6\,mm 
diameter was sandwiched by thin gold foils of the same dimensions. 
At 25\,keV neutron
energy measurements were carried out with calcium carbonat sample masses 
between
7 and 70\,mg. No significant effect from multiple neutron scattering was
observed. A sample of the $^{48}$CaCO$_3$ powder was also heated to 
300$^{\rm o}$\,C for 3 hours. A weight loss of only 0.45\,\%  may be ascribed to 
absorbed water. In Fig.~\ref{ff3}
the accumulated
$\gamma$-ray intensity from one of the $^{48}$Ca activations is shown. The
$\gamma$-line is well isolated on a low level of background counts.

As it was very difficult to press the $^{48}$CaCO$_3$ powder to stable
self supporting tablets, the material was in most cases filled into 
cylindrical polyethylene or teflon containers. When we succeeded to press
a somewhat stable tablet it was put into a thin Al-foil. No measurable
effects due to the containment were detected. As our measurements are
significantly below an earlier measurement \cite{kae85} especially at
25\,keV neutron energy all these checks were performed. To be sure that
the enrichment is correct an additional mass spectrometric analysis
was carried out by the "Wiederaufarbeitungsanlage Karlsruhe (WAK)" which 
confirmed the original certificate. As the detector showed also radiation
damage in the last runs, the efficiency was rechecked with radioactive 
samples. Finally, to the sandwich of gold foils a sandwich of Ag-foils
was added in the run at 25\,keV with the 7\,mg $^{48}$CaCO$_3$ sample. If we 
relate in this run the $^{48}$Ca cross-section to the previously measured
capture cross sections of $^{107,109}$Ag \cite{bee94}, we obtain
(855$\pm$150$)\,\mu$barn which is indeed higher by 14\,\% but within quoted
uncertainties agrees with our measurements related to the $^{197}$Au
capture cross-section. 
 
The following systematic uncertainties were combined by quadratic error
propagation; Au standard cross section: 1.5-3\,\%, Ge-detector efficiency:
8\,\%, $\gamma$-ray intensity per decay: 1.1\,\% for the $^{49}$Ca and 
0.1\,\%
for the $^{198}$Au decay, divergence of neutron beam: 2--7\,\%, factor
$f_{\rm b}$: 1.5\,\%, sample weight: $<$0.5\,\%, and other systematic errors: 
3\,\%. 
\section{Calculations and results}\label{s5}
Neutron capture on several calcium isotopes has been analyzed recently
by Krausmann {\it et al.} \cite{kra95} using the direct-capture (DC) model together with
folding potentials. The DC formalism was developed by 
Kim {\it et al.}~\cite{kim87}, and can also be found in~\cite{kra95,mohr93}.
Here we only repeat some relations
which are important for the subsequent calculations.

The DC cross-section for the capture to a final state $i$ is given by
\begin{eqnarray}
\sigma_i^{{\rm DC}} 
  & = & 
   \int {\rm d}\Omega \, \frac{{\rm d}\sigma^{{\rm DC}}}{{\rm d}
   \Omega_\gamma} \nonumber \\
  & = & 
   \int {\rm d}\Omega\,2\,\left( \frac{e^2}
  {\hbar\,{\rm c}} \right) \left( \frac{\mu {\rm c}^2}{\hbar\,{\rm c}} \right)
  \left( \frac{k_\gamma}{k_a} \right)^3 \frac{1}{2\,I_A + 1}\,
  \frac{1}{2\,S_a + 1} \sum_{M_A\,M_a\,M_B,\,\sigma}
  \mid T_{M_A\,M_a\,M_B,\,\sigma} \mid^2 \quad. 
\label{eq:DC}
\end{eqnarray}
The quantities $I_A$, $I_B$ and $S_a$ ($M_A, M_B$ and $M_a$)
are the spins (magnetic quantum numbers) of the target nucleus $A$, 
residual nucleus $B$ and projectile $a$, respectively. 
The reduced mass in the entrance channel is given by $\mu$. The
polarization $\sigma$ of the electromagnetic radiation can be $\pm 1$. The wave
number in the entrance channel and for the emitted radiation is given 
by $k_a$ and $k_\gamma$, respectively. 
The total capture cross-section is given by the sum over all
DC cross-sections for each final state $i$,
multiplied by the spectroscopic factor $C^2 S$
which is a measure for the probability of finding
$^{49}$Ca in a ($^{48}$Ca $\otimes$ n) single-particle configuration
\begin{equation}
\sigma^{\rm{th}} = \sum_i C^2_i S_i \, \sigma_i^{\rm{DC}} \quad .
\label{eq:C2S}
\end{equation}

The transition matrices $T = T^{\rm E1} + T^{\rm E2} + T^{\rm M1}$ depend on
the overlap integrals
\begin{equation}
I^{{\rm E}{\cal L}/{\rm M}{\cal L}}_{l_b\,j_b\,I_B;l_a\,j_a} = 
   \int\,{\rm d}r\,U_{l_b\,j_b\,I_B}(r)
   {\cal O}^{{\rm E}{\cal L}/{\rm M}{\cal L}}(r)\,\chi_{l_a\,j_a}(r) \quad
\label{ac}
\end{equation}
The radial part of the bound state wave function in the exit channel and the 
scattering
wave function in the entrance channel is given by $U_{l_b\,j_b\,I_B}(r)$ and
$\chi_{l_a\,j_a}(r)$, respectively. The radial parts of the electromagnetic
multipole operators are well-known (e.g., \cite{kra95}). The calculation of
the DC cross-sections has been performed using the code TEDCA~\cite{TEDCA}.

The most important ingredients in the potential models are the wave functions 
for the scattering and bound states in the entrance and exit channels. 
For the calculation of the wave functions we use real Saxon-Woods (SW) 
potentials as well as real folding potentials which are given by
\begin{equation}
V(R) = 
  \lambda\,V_{\rm F}(R) 
  = 
  \lambda\,\int\int \rho_a({\bf r}_1)\rho_A({\bf r}_2)\,
  v_{\rm eff}\,(E,\rho_a,\rho_A,s)\,{\rm d}{\bf r}_1{\rm d}{\bf r}_2 
  \label{ad}
\end{equation}
with $\lambda$ being a potential strength parameter close
to unity, and $s = |{\bf R} + {\bf r}_2 - {\bf r}_1|$,
where $R$ is the separation of the centers of mass of the
projectile and the target nucleus.
The density of $^{48}$Ca has been derived from the measured
charge distribution \cite{vri87}, and the effective nucleon-nucleon
interaction $v_{\rm eff}$
has been taken in the DDM3Y parametrization \cite{kobos84}.
The resulting folding potential has a volume integral per interacting
nucleon pair $J_R = 446.63\,{\rm{MeV\,fm^3}}$ ($\lambda = 1$)
and an rms-radius $r_{\rm rms} = 4.038\,{\rm{fm}}$.
Details about the folding procedure can be found for instance in 
\cite{abele93}, the folding potential has been calculated by using the
code DFOLD \cite{DFOLD}. The imaginary part of the potential 
is very small because of the small flux into reaction channels \cite{kra95}
and can been neglected in our case.

\begin{figure}[t]
\centerline{\psfig{file=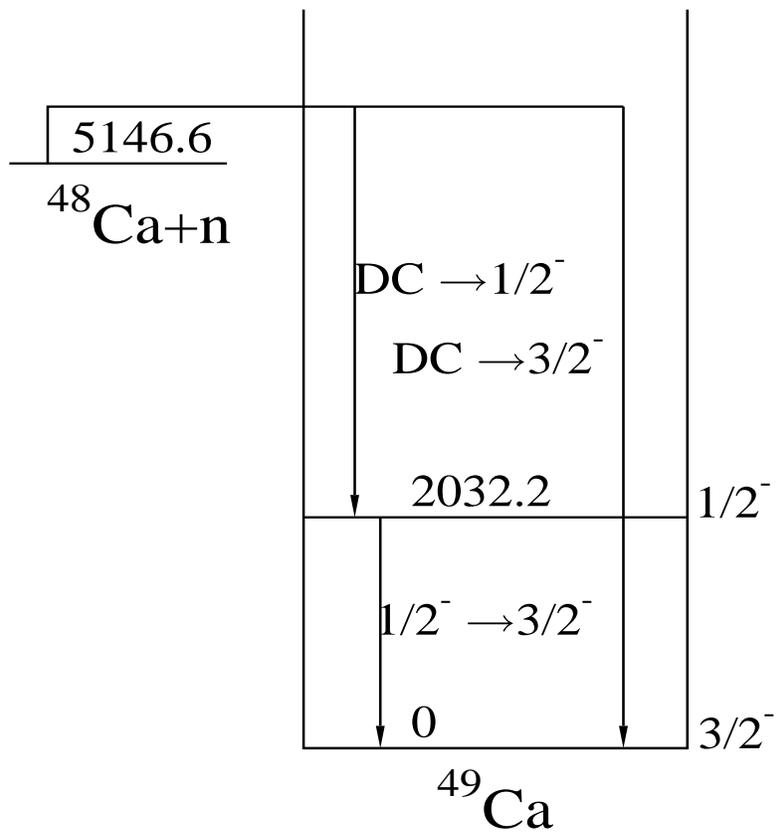,width=18cm,height=16cm}}
\vspace{1cm}
\caption{\label{ff4} Partial level scheme of $^{49}$Ca}
\end{figure}

Two important results have been obtained in the work of 
Krausmann {\it et al.} \cite{kra95}: first, the $^{48}$Ca(n,$\gamma$)$^{49}$Ca
reaction can be well described in terms of a direct-reaction mechanism,
and second, only the E1-transitions from the incoming s-wave to the
ground state of $^{49}$Ca ($3/2^-$) and to the first excited state at
$E^* = 2023\,{\rm{keV}}$ ($1/2^-$) play a significant role in the
thermal and thermonuclear energy range (Fig.~\ref{ff4}).

\begin{figure}[t]
\centerline{\psfig{file=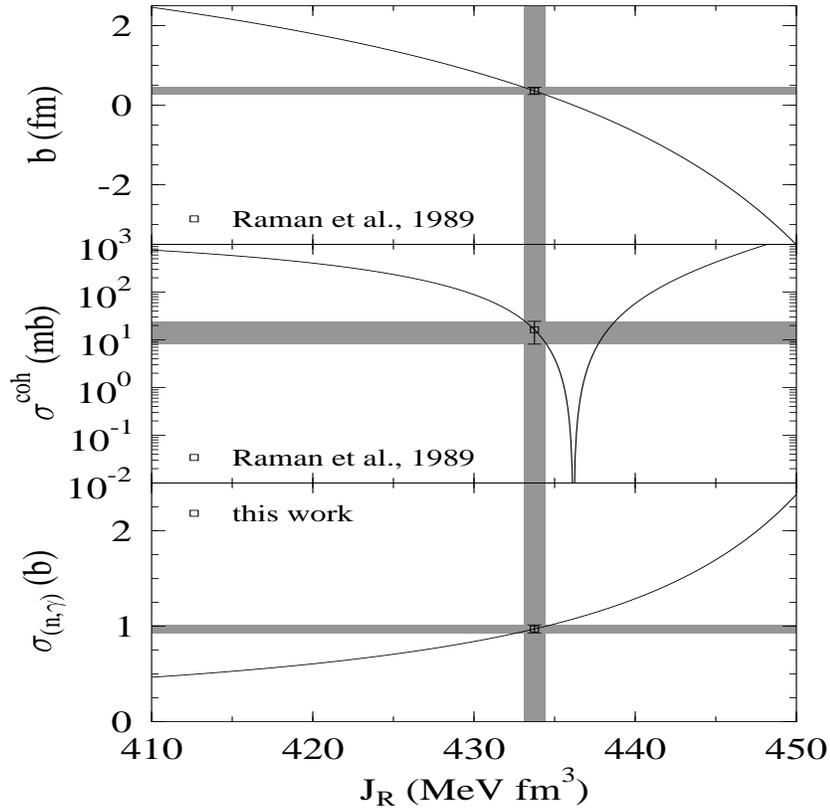,width=18cm,height=16cm}}
\vspace{1cm}
\caption{\label{ff5} 
Neutron scattering length $b$,
coherent elastic scattering cross section $\sigma^{\rm coh}$, and
thermal capture cross-section (and their uncertainties, gray shaded)
in dependence of the volume integral of the folding potential: 
the potential strength can be adjusted very accurately because of
the very small scattering length which is close to zero.}
\end{figure}

The neutron scattering length on $^{48}$Ca has been measured by 
Raman {\it et al.}~\cite{raman89}: $b = (0.36 \pm 0.09)\,{\rm{fm}}$.
As can be seen from Fig.~\ref{ff5} (upper part), the optical potential
can be adjusted very accurately to the scattering length or to
the coherent scattering cross-section (which is given by
$\sigma^{\rm coh} = 4 \pi b^2 = (16.3 \pm 8.1)\,{\rm{mb}}$),
because even
small changes in the strength of the potential result in drastic changes
of the scattering length. For the folding potential we obtain
$\lambda = 0.9712 \pm 0.0015$, and $J_{\rm R} = (433.75 \pm 0.65)$\,MeV\,fm$^3$, 
for a SW potential
with standard geometry ($r_0 = 1.25$\,fm, $a = 0.65$\,fm)
we obtain $V_0 = (47.74 \pm 0.08)$\,MeV, and
$J_{\rm R} = (469.57 \pm 0.79)$\,MeV\,fm$^3$.
The folding potential has only one parameter $\lambda$ which can be
adjusted directly to the scattering length; the geometry of this potential
is fixed by the folding procedure. This is a clear advantage compared
to the usual SW potential, where one has to choose reasonable values
for the radius parameter $r_0$ and the surface diffuseness $a$.

\begin{figure}[t]
\centerline{\psfig{file=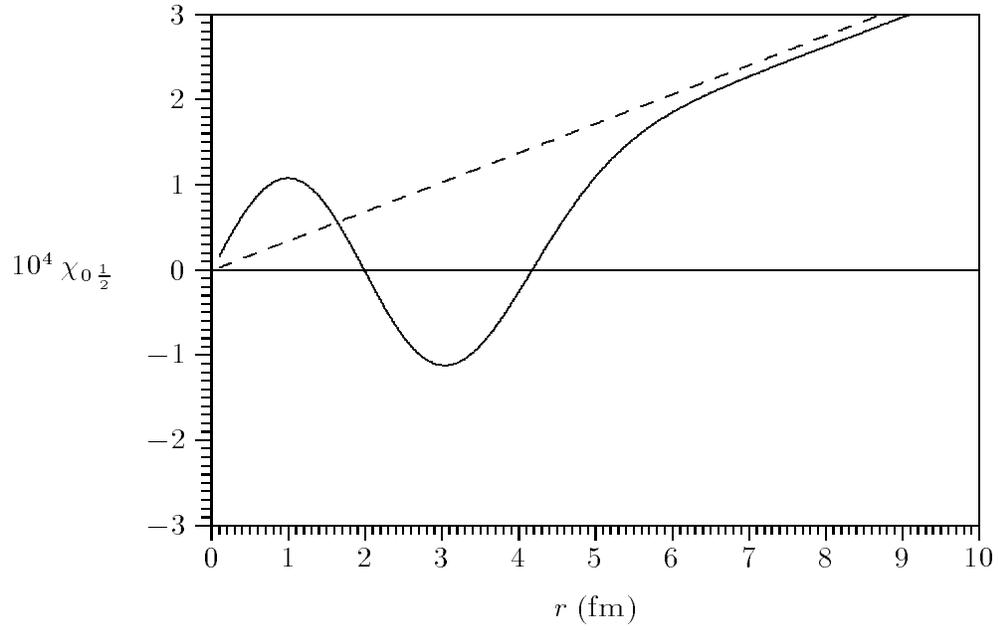,width=18cm,height=16cm}}
\vspace{1cm}
\caption{\label{ff6} Real part of the
($^{48}$Ca+n)-scattering wave function $\chi_{\ell=0, j=1/2}$ using a folding potential fitted
to the experimental scattering cross section (solid curve),
and for a vanishing potential (dashed curve)}
\end{figure}

From Fig.~\ref{ff5} (central part) one notices that the value of the coherent
(n+$^{48}$Ca)-scattering
cross-section almost disappears for a volume integral near
$J_{\rm R}$ =  436\,MeV\,fm$^3$.
In fact, the experimental coherent elastic cross section for
(n+$^{48}$Ca)-scattering (horizontal shaded
area in the central part of Fig.~\ref{ff5}) has by far the lowest
value for all known target nuclei~\cite{sea92}.
This is due to the so-called nuclear Ramsauer
effect~\cite{law53,pet62}, which is analogous
to the already long--known electron--atomic effect~\cite{ram21}. The condition for a
minimum in the elastic cross section is that the phase-shift difference
of the neutron paths traversing the
nucleus and going around is
$\Delta = n \pi$ with $n =$ even~\cite{pet62}. For low energies
the phase shift for going outside the nucleus can be neglected so
that the above expression reduces to
$\delta = \Delta/2 = m \pi = n \pi/2$ with $m =$ integer,
where $\delta$ is now the normal scattering phase shift.
As can seen from the solid curve in Fig.~\ref{ff6} for the phase shift in 
(n+$^{48}$Ca)--scattering $m = n/2 = 2$. Almost the same
scattering wave function in the nuclear
exterior is obtained for a vanishing potential (broken
curve in Fig.~\ref{ff6}).
That means
that near zero neutron energy almost
exactly two neutron--wavelenghts ($m=2$, $n=4$) fit inside the $^{48}$Ca nucleus, giving a
pronounced minimum in the cross section.

This neutron--nuclear situation contrasts with
the electron--atomic Ramsauer effect, where only
one minimum ($m=1$, $n=2$) is possible.
Elastic neutron scattering on $^{48}$Ca is an excellent
example for the nuclear Ramsauer effect, because neutron-induced
low-energy absorption
into other channels is (compared with other nuclei) exceptionally
low~\cite{sea92}. This means that for elastic neutron scattering on $^{48}$Ca
also the elastic shadow scattering is very small. Both effects
(nuclear Ramsauer effect, almost no absorption into other channels)
lead to the extremely small cross section in elastic low-energy
neutron scattering on $^{48}$Ca.

The bound state wave functions are calculated using the same folding-potential
shape as for the scattering waves. Only the potential strengths have been
adjusted to reproduce the binding energies of the $3/2^-$ and $1/2^-$
bound states. The resulting parameters are listed in 
Table \ref{tab:bound}.

According to Eq.~(\ref{eq:C2S}) spectroscopic factors $C^2 S$
of the two bound states
can be extracted from the ratio of the measured thermal capture cross-section
of the two final states and the calculated thermal DC cross-section.
This can again be done with high accuracy because the
calculated DC cross-section is not very sensitive to small changes 
of the strength of the optical potential (see lower part of Fig.~\ref{ff5}).
The resulting spectroscopic factors (see Table \ref{tab:bound})
depend only weakly on the chosen
potential (folding potential or Saxon-Woods potential).
The spectroscopic factors are somewhat smaller than values
determined in a recent (d,p)-experiment but both values agree within their
uncertainties which are smaller in this determination of $C^2 S$
compared to the extraction of $C^2 S$ from the analysis of a (d,p)-experiment
(see Table~\ref{tab:bound}).

\begin{figure}[t]
\centerline{\psfig{file=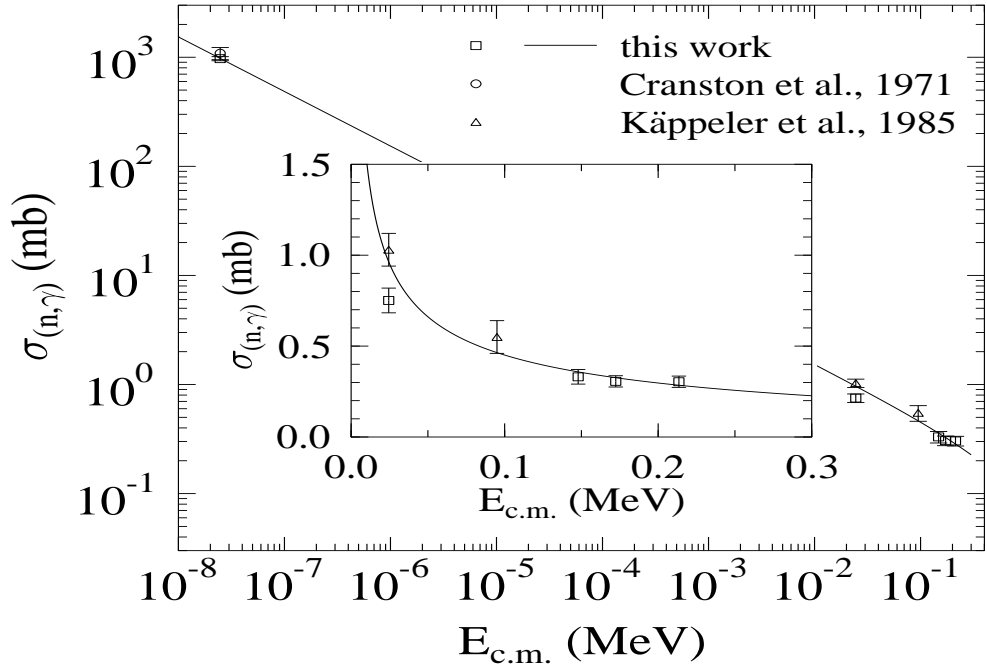,width=18cm,height=16cm}}
\vspace{1cm}
\caption{\label{ff7} Comparison of the direct-capture cross-section for
$^{48}$Ca(n,$\gamma$)$^{49}$Ca at thermal and thermonuclear energies
with the experimental data}
\end{figure}

For the calculation of the thermonuclear capture cross-section
no parameters have to be adjusted to the experimental data.
The optical potentials, bound state wave functions and spectroscopic
factors have been determined from the scattering length and the
thermal capture cross-section. In Fig.~\ref{ff7} the result of our calculation
is compared to the experimental data available in literature
\cite{cra71,kae85} and to our
new experimental data. Good agreement between the DC calculation and
the experimental data is obtained.

Only our experimental value
at a neutron energy of $E_{\rm n} = 25\,{\rm{keV}}$ is about 25\,\%
lower compared to the calculated direct-capture cross-section
as well as the one obtained from extrapolating the thermal cross
section with an 1/v-behavior.
This difference could be explained through a destructive
interference of the direct part with a resonance having the parameters
$E_{\rm res} = 1.45$\,keV, $\Gamma_{\rm n} = 0.33$\,keV,
and $\Gamma_{\gamma} = 0.22$\,eV.
However, in the neutron--capture data taken at ORELA
some years ago~\cite{car85,car87} such a strong resonance peak should have been
observed~\cite{mac96}. The two small resonances at 19.3\,keV and 106.9\,keV
that have been observed are much weaker than the proposed resonance.

In Table~\ref{tab:spec} the neutron spectroscopic factors of $^{49}$Ca determined from
shell-model calculations are compared with the values
extracted from the experimental cross sections for
$^{48}$Ca(d,p)$^{49}$Ca and $^{48}$Ca(n,$\gamma$)$^{49}$Ca.
This comparison of the experimental data with the calculated values
is especially important as
they form the basis for the extrapolation of the calculations
of neutron capture on neutron--rich unstable target nuclei
in the mass number range $A = 36$ to $A = 66$, where
experimental data is scarcely or not at all available. However, the knowledge
of these cross sections is an essential part in the explanation
of the isotopic anomalies of Ca--Al--rich inclusions of certain 
primitive meteorites~\cite{nie84,har93}. The cross sections and reaction
rates must be calculated via the direct reaction model using the
input data of nuclear structure models, for instance the shell
model. The spectroscopic factors calculated in the shell
model have been obtained using the code OXBASH \cite{Bro84}. We have employed the
interaction FPD6 of Richter et al.~\cite{Ric91} which
assumes an inert $^{40}$Ca--core and the remaining nucleons in the
full fp-shell. As can be seen from Table~\ref{tab:spec} the spectroscopic
factors agree well for the ground state, the first $1/2^-$ and the
second $5/2^-$ state. These states are almost pure single particle
states. The first $5/2-$ and the second $1/2^-$ state cannot
be reproduced with this interaction.

We have determined the thermonuclear-reaction-rate factor
$N_{A}\left<\sigma v \right>$~\cite{fow67}. Since the cross
section follows an $1/v$--law up to at least 220\,keV we obtain a
constant reaction--rate factor
\begin{equation}\label{3}
N_{A}\left<\sigma v \right> =
1.28 \times 10^5 \,{\rm cm}^3\,{\rm mole}^{-1}\,{\rm s}^{-1} \quad .
\end{equation}

\section{Summary}\label{s6}
We have measured the thermal and thermonuclear (25, 151, 176, and 218\,keV)
$^{48}$Ca(n,$\gamma$) cross-sections. From these data we calculated
the thermonuclear-reaction-rate factor. We found the thermal capture
cross-section 10\,\% lower than the previous value of Cranston and
White \cite{cra71}. In the analysis our thermal cross section was
found to be consistent with our thermonuclear values assuming a 1/v-behavior,
except the 25\,keV value which was lower by 25\,\%.
The previous thermonuclear values of K\"appeler et al.~\cite{kae85}
are satisfactory compared with the fit to our experimental data.
As anticipated by previous experimental results there is only negligible
influence of the resonance part of the cross section up to at least
220\,keV. Therefore, the neutron capture by $^{48}$Ca can be well described
in this case by using a direct-capture model. 
\section*{Acknowledgements}
We are grateful to Richard L. Macklin and Jack Harvey for giving us additional valueable
information of their $^{48}$Ca time--of--flight measurements.
From the IRMM, Geel, we thank C. Van der Vorst and E. Macavero for their
help in the thermal measurement at the reactor BR1 in Mol. The support of
the BR1 reactor staff is also gratefully acknowledged.
We would like to thank the technician of the Van de Graaff G. Rupp and the
Van de Graaff staff members E.P. Knaetsch, D. Roller, and W. Seith for 
their help and support of the experiment especially in the
preparation of the metallic Li--targets.
We thank the  Fonds zur F\"orderung der
wissenschaftlichen Forschung in \"Osterreich (project S7307--AST),
the \"Ostereichische Nationalbank (project 5054) and the Deutsche
Forschungsgemeinschaft (DFG) (project Mo739/1--1)
for their support.

\begin{table}
\caption{\label{tt1} Sample characteristics and decay properties of the
product nuclei $^{49}$Ca and $^{198}$Au}
\begin{center}
\begin{tabular}{ccccccc}
Isotope&Chemical &Isotopic composition & Reaction & $T_{1/2}$ & $E_\gamma$ &
Intensity per decay\\
  &form   &  (\%)   &   &   &  (keV)    & (\%) \\
\hline
$^{48}$Ca&CaCO$_3$&20.42(40), 0.24(42), 0.060(43), 1.48(44),&
$^{48}$Ca(n,$\gamma$)$^{49}$Ca&(8.716$\pm$0.011)\,min& 3084.54 &92.1$\pm$1.0 
\\ && 0.01(46), 77.87$\pm$1.90(48)   &&&&\\
$^{197}$Au&metallic& 100&$^{197}$Au(n,$\gamma$)$^{198}$Au&2.69\,d& 412&
95.50$\pm$0.096\\
\end{tabular}
\end{center}
\end{table}

\begin{table}
\unitlength1cm
\caption{\label{th2} Sample weight and experimental $^{48}$Ca
capture at thermal energy}
\begin{center}
\begin{tabular}{ddddddd}
Sample & Weight&Isotope&\multicolumn{3}{c}{Primary $^{49}$Ca Transitions}&
$\sigma_{\rm thermal}$ \\
\cline{4-6}
& (mg) &&$J^{\pi}$&$E_{\gamma}$&$f_{\gamma}$ & (mbarn)  \\
&&& & (keV) & (\%) & \\
\hline
        &        &         &3/2$^-$&5142 &74$\pm$3  &\\
CaCO$_3$& 108.630&$^{48}$Ca&       &     &          &982$\pm$46\\
        &        &         &1/2$^-$&3121 &23$\pm$1    &\\
\end{tabular}
\end{center}
\end{table}

\begin{table}
\unitlength1cm
\caption{\label{tt2} Sample weights and experimental $^{48}$Ca
capture cross-sections at thermonuclear energies}
\begin{center}
\begin{tabular}{dddddddd}
Mean neutron  & Mass of Au &  Mass of  & Mass of Au &
Irradiation &
$\sigma$ &\multicolumn{2}{c}{Uncertainty}\\
energy& front side& CaCO$_3$& back side& time  & ($\mu$barn) &
statistical&total\\
(keV)   &   (mg)  &  (mg) & (mg)  & (d) &  &(\%)&(\%)\\
\hline
25    &  15.577   &  18.28 & 15.583&   0.98  &  721&1.4& 10\\
      &  15.745   &  10.064& 15.755&   0.94  &  719&0.9& 10\\
      &  15.390   &  21.353& 15.352&   0.80  &  767&1.4& 9 \\
      &  16.010   &  8.6025& 15.950&   0.86  &  817&1.5& 10\\
      &  16.08   &   11.243& 16.115&   0.97  &  781&3.8& 10\\
      &  15.915  &    7.017& 15.782&   0.91  &  733&3.4& 10\\
\cline{6-8}
\multicolumn{5}{r}{Average}&                    751$\pm$68&&\\
\hline
$151 \pm 15$  &15.657 & 49.914&15.673&1.77  &   331 &2.1&12\\
\cline{6-8}
\multicolumn{5}{r}{Average}&                    331$\pm$40&&\\
\hline
$176 \pm 20$  &15.553 & 70.58& 15.570& 3.70  &  303&1.5&11\\
              &15.186 & 67.22& 15.248& 3.90  &  310&2.0&11\\
\cline{6-8}
\multicolumn{5}{r}{Average}&                    306$\pm$31&&\\
\hline
$218 \pm 23$  &16.740 &58.150&16.783&  5.66  &  318&2.1&10\\
      &        15.482 &68.790&15.532&  4.07  &  294&1.3&10\\
\cline{6-8}
\multicolumn{5}{r}{Average}&                   304$\pm$31&&\\
\end{tabular}
\end{center}
\end{table}

\begin{table}
\caption{\label{tab:bound} 
        Final states of the reaction $^{48}$Ca(n,$\gamma$)$^{49}$Ca:
        $J^\pi$, excitation energies, potential parameters, and spectroscopic
        factors.}
\begin{center}
\begin{tabular}{ccccccccc}
$J^\pi$ & $E^*$ & $\lambda$ & $J_{\rm R}^{\rm fold}$ & $V_0$ & $J_{\rm R}^{\rm SW}$
        & $C^2 S^{\rm fold}$ & $C^2 S^{\rm SW}$ & $C^2 S_{\rm (d,p)}$ \\
        & (keV) &           & (MeV\,fm$^3$) & (MeV) & (MeV fm$^3$)
        & & & (Ref.~[5]) \\
\hline
$3/2^-$ & 0.0    & 0.9892 & 441.82 & 49.622 & 488.06 
        & $0.72 \pm 0.04$ & $0.73 \pm 0.04$ & $0.84 \pm 0.12$ \\
$1/2^-$ & 2023.2 & 0.9111 & 406.92 & 45.639 & 448.89 
        & $0.86 \pm 0.05$ & $0.85 \pm 0.05$ & $0.91 \pm 0.15$ \\
\end{tabular}
\end{center}
\end{table}

\begin{table}
\caption{\label{tab:spec}
                Spins, excitation energies, Q--values and spectroscopic factors of
                levels in $^{49}$Ca extracted from (n,$\gamma$)-data (this work),
                (d,p)-data~[5], and the shell model (SM)}
\begin{tabular}{rddddd}
$J^\pi$ & $E^*$ [MeV] & $Q$ [MeV] & $S_{({\rm n,}\gamma)}$ &
$S_{\rm (d,p)}$ & $S_{\rm SM}$\\ \hline
$3/2^-$ & 0.000 & 5.142 & 0.72 & 0.84 & 0.91 \\
$1/2^-$ & 2.023 & 3.121 & 0.86 & 0.91 & 0.94 \\
$5/2^-$ & 3.586 & 1.556 &  --  & 0.11 &  --  \\
$5/2^-$ & 3.993 & 1.149 &  --  & 0.84 & 0.93 \\
$3/2^-$ & 4.069 & 1.073 &  --  & 0.13 & 0.01 \\
$1/2^-$ & 4.261 & 0.881 &  --  & 0.12 &  --  \\
\end{tabular}
\end{table}

\newpage

\begin{figure}[t]
\centerline{\psfig{file=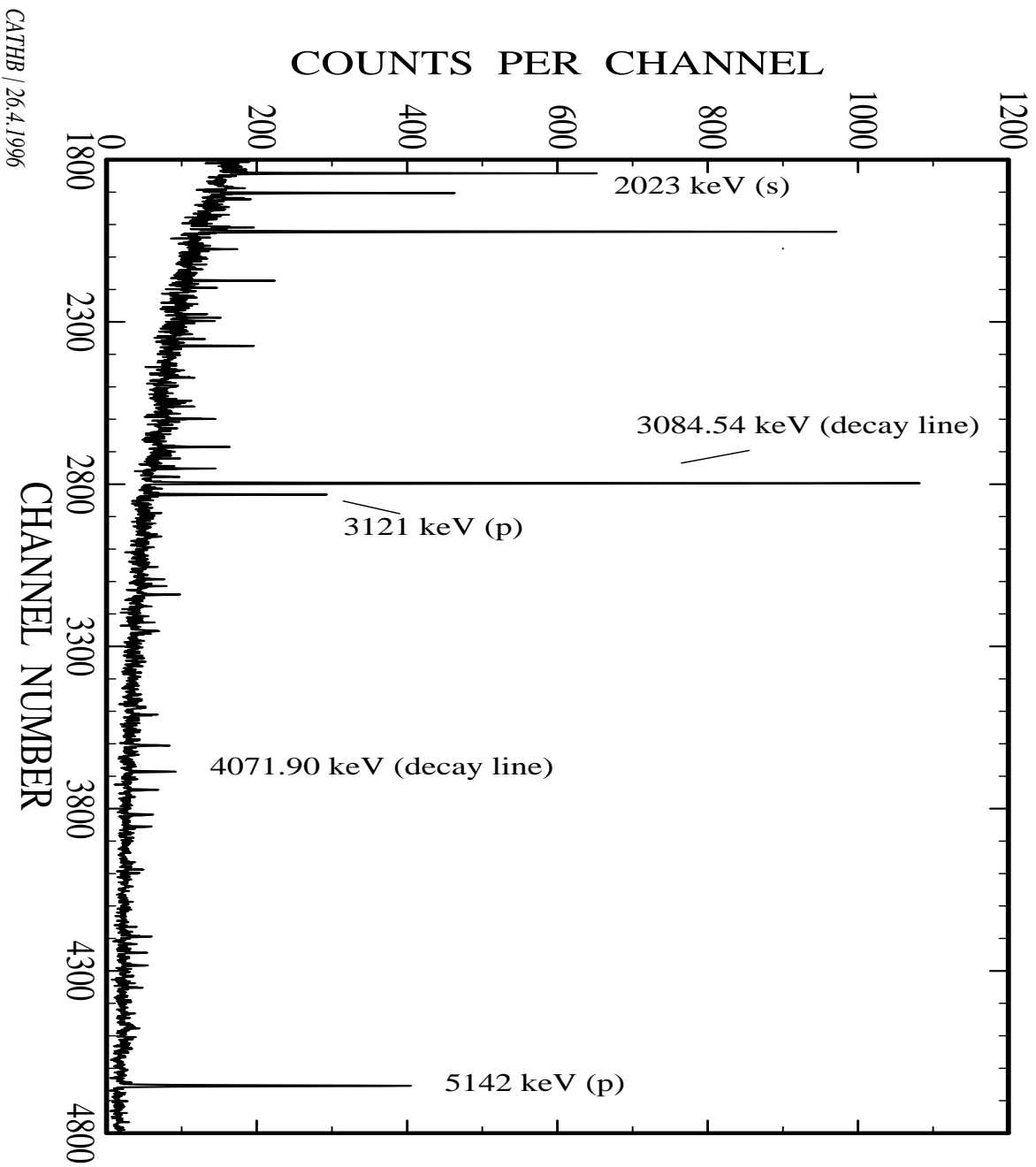,width=18cm,height=16cm,angle=270}}
\vspace{1cm}
\caption{\label{ff2} Accumulated intensities of the prompt $^{49}$Ca and 
$\beta$--delayed
$^{49}$Sc $\gamma$-lines
from the thermal experiment}
\end{figure}

\end{document}